# Hadron Colliders

*M. Zerlauth and O. Brüning*
CERN – European Organisation of Nuclear Research, Geneva, Switzerland

**Abstract**
In this paper we will provide an overview of the hadron colliders built to date and the design and operational challenges that each of these machines has faced.
Many of these are inherent to the ongoing effort to optimise the instantaneous and integrated luminosity of the machines, which inevitably lead to many technological challenges that must be met and overcome. We will summarise how these challenges have been successfully met in the past and present machines and outline the role they could play in ambitious future accelerator projects such as the HL-LHC upgrade and the FCC project.

**Keywords**
Luminosity optimisation, beam lifetime, operational challenges, performance challenges, high luminosity, high intensity.

## 1 Performance Challenges in High Luminosity, High Intensity circular colliders

Present and future hadron colliders are striving for every higher performance in terms of collision energy and beam intensity, to increase the instantaneous and integrated luminosities that are the driver of the discovery potential of these exploration machines. The design and operation of such high intensity, high brightness machines however come with several performance challenges, linked to

– the need to increase data volume and collision energy,

– luminosity optimization, beam lifetime and operational challenges,

– material and technology challenges.

### 1.1 Increase of data volume and collision energy

Colliders are built to accelerate either leptons or hadrons. Leptons are true elementary particles, featuring a very well-defined centre of mass energy, being therefore the ideal choice for precision measurements. On the downside, they are very light particles (g >>1) which will exhibit strong synchrotron radiation and therefore intrinsically limit the energy reach of the collider.
Hadrons enable multi particle collisions and offer a range of centre of mass (CM) energies, therefore typically being the particle of choice for discovery machines at the price of high backgrounds (i.e. generating a lot of known particles) for the experiments. Their much higher weight effectively supresses synchrotron radiation but will require much stronger magnetic fields for their guidance along the circular trajectory. The development and industrialisation of superconducting magnet technologies represent therefore the driving factor for the energy reach of today's hadron machines.

The key performance measures of hadron colliders are the collision energy, instantaneous luminosity (the rate of events in the detector) and the integrated luminosity (total number of events recorded by the detector). CERNs Large Hadron Collider (LHC) is todays flagship of the High Energy Physics community, delivering proton-proton collisions at 6.8 TeV per beam, instantaneous luminosities exceeding $2\times10^{34}$ cm$^{-2}$sec$^{-1}$ in the two high luminosity experiments of ATLAS and CMS and has allowed to collect already close to 500 fb$^{-1}$ by the middle of 2025, promising to reach 550 fb$^{-1}$ by the end of the nominal operational period of the LHC in June 2026.

To enable new discoveries in particle collisions, the CM energy must be above the production threshold of the sought particle (e.g. for the Z above 45 GeV, for the W above 80 GeV, for the Higgs above 125 GeV), fuelling the desire to reach the highest possible CM energies in future hadron colliders. Optimising the performance of hadron colliders however requires considering two competing effects, namely:

1. Production rate: the cross section decreases with the particle mass (see Fig. 1). Pushing the discovery frontier requires therefore also an increase in luminosity with every increase in collision energy [1].

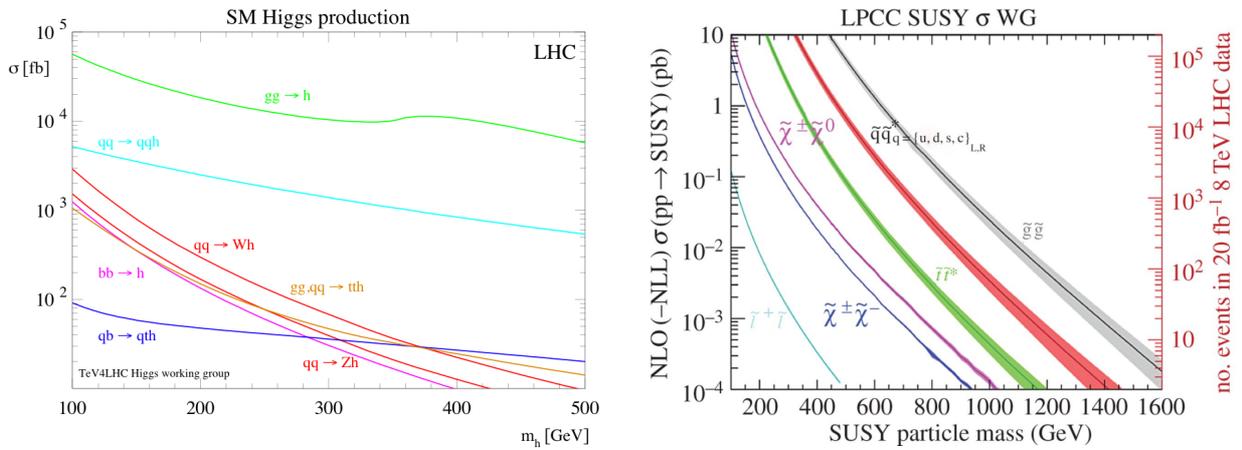

**Fig. 1:** Production rates as a function of particle mass.

2. Beam Burn off: the total cross section increases with CM energy (see Fig. 2). Since beam lifetime decreases with increasing luminosity and energy, the resulting shorter physics fills imply a reduced machine efficiency and therefore a reduced integrated luminosity. Reduced beam lifetime also implies an increased background and an increased radiation load, both in the detectors and the machine.

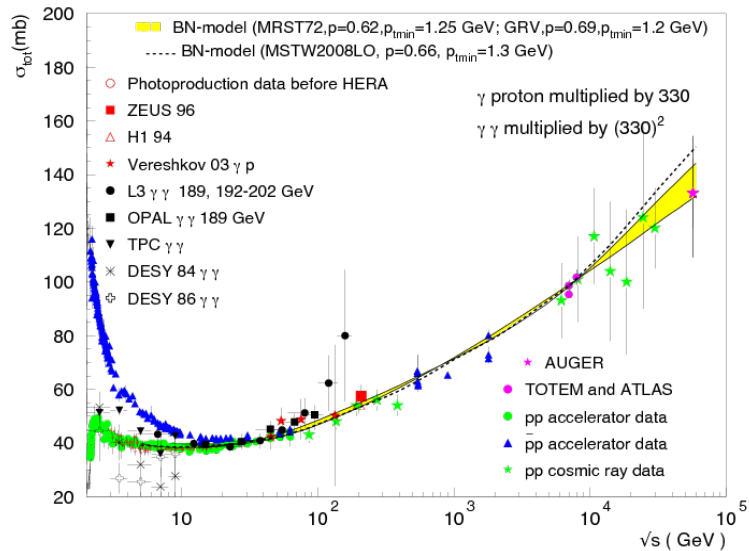

**Fig. 2:** Total cross-section as function of collision energy.

## 1.2 Luminosity optimization and beam lifetime

To maximise the luminosity of a hadron collider, the designs will aim for the largest number of bunches with high bunch intensity, and a small beam size ($\beta^*$) at the point of collision. The achievable bunch intensities and beam emittance are typically given and/or limited by the performance of the injector complex, the minimum beam size at the IP will be driven by the available magnet aperture in the final focusing regions, and the desired number of bunches will drive the choice of crossing-angle to avoid parasitic encounters and long-range beam-beam effects. When having to choose a larger crossing angle due to the higher bunch intensities and number of bunches, the compensation of the geometric loss factor (F) becomes important, which has been achieved in the High-Luminosity LHC using so-called crab-cavities.

While not a direct design parameter of the collider, machine efficiency and the underlying need to minimize the number of faults and turnaround time has proven a vital performance driver of present colliders such as the LHC. Considering that the effective beam lifetime is inversely proportional to the achieved luminosity, requires that the average fill length (i.e. the available time for data taking for physics) is significantly longer than the time required for preparing a new fill for physics to ensure the efficient operation of such complex machines. Operation at the luminosity frontier is therefore making use of luminosity levelling techniques, allowing to keep the luminosity artificially lower during operation, but in turn sustain these levelled luminosity levels for longer durations in each physics fill. At the same time, this helps to cope with the total number of events per time interval and the event pileup in the detectors. The event rate and pile-up, as well as quench protection, heat load and radiation in accelerator equipment due to debris particles escaping from the collisions points are key drivers of luminosity limitations in today's hadron colliders.

Several luminosity levelling techniques have been developed and are today already in operational use in CERNs Large Hadron Collider, listed here below with increasing level of complexity for their implementation:

– levelling by beam offset / separation,
– levelling by crossing angle, and
– levelling by $\beta^*$ (=beam size at the IP).

Experience at previous Hadron Beam Colliders has shown that the beam lifetime and losses are very sensitive to the transverse profile matching and overlap of the colliding beams at the interaction point. To limit the effects of non-linear fields from long-range beam-beam interaction, the push for larger bunch intensities will require operation at larger separations up to 10-12 $\sigma$, requiring larger crossing-angles at the interaction point. For the HL-LHC configuration, the crossing angle must be increased to 500 µrad compared to the nominal LHC design of 285 µrad. This larger crossing angle considerable reduces the effective cross-section of the two colliding bunches as shown in Fig. 3. While this effect is not yet very dominant for the LHC design values, the much smaller $\beta^*$ in the HL-LHC era will reduce the effective cross-section by some 70 % compared to the case of a perfect overlap. To compensate for this loss of effective cross-sections, crab-cavities have been introduced in the baseline of the HL-LHC upgrade. While not required in the beginning of the fill (where crossing angle and $\beta^*$ levelling techniques will be used to level at the design luminosity of $5\times10^{34}$ cm$^{-2}$sec$^{-1}$), the crab cavities will be switched on towards the end of the fill to prolong the levelling time at constant nominal luminosity. Like in the HL-LHC, crab cavities are also a fundamental part of the Electron-Ion Collider (EIC) project to compensate for the large crossing angle of 25 mrad. Despite these efforts, the luminosity frontier today is still with SuperKEKB, which after having achieved $4.71\times10^{34}$ cm$^{-2}$sec$^{-1}$ in June 2022 is today on an upgrade path towards the ultimate design luminosity of $6.5\times10^{35}$ cm$^{-2}$sec$^{-1}$.

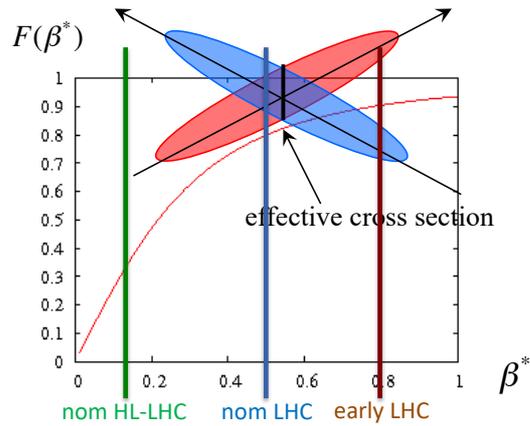

**Fig. 3:** Geometric loss factor as a function of $\beta^*$.

## 1.3 Material and technology challenges

The push for higher beam intensities and collision energies is also posing increasing challenges to the materials and technologies used for the construction of accelerator and beam line equipment. Three main challenges can be identified to this end:

- damage potential and machine protection (due to the increasing beam power and stored energies),
- quench protection, heat load and radiation effects (due to debris from the collision points), and
- magnet technology and accelerator size (due to the increasing beam energy).

### 1.3.1 Damage potential and machine protection

With increasing beam energy and beam intensities, the stored energies to be handled in today's hadron colliders (both in terms of beam energy as well as the magnetic stored energy) exceed the damage thresholds by several orders of magnitude (see Fig. 4).

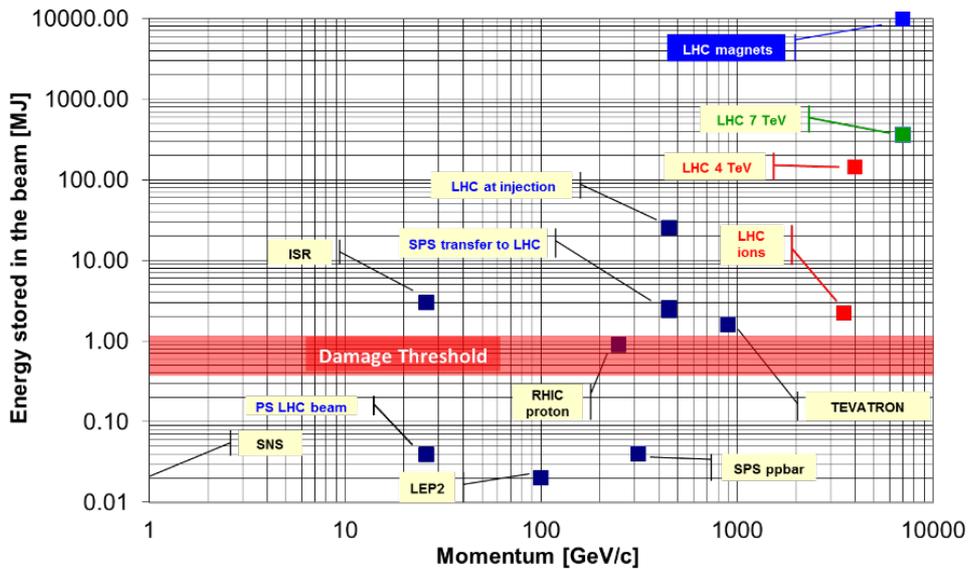

**Fig. 4:** Stored beam and magnet energies in recent lepton and hadron colliders.

RHIC, the TEVATRON and even a full SPS beam at extraction already exceed the damage threshold in case of a localized impact of the full beam energy. Considerations for machine protection do however need to account for additional relevant parameters beyond the momentum of the particle and the total energy stored in the beam, such as the particle type, the beam power (CW or pulsed), the beam size and the time structure of the beam. The uncontrolled release of such stored energies can result in considerable damage as has shown the release of approximately 600 MJ of stored magnetic energy in the 2008 incident at the LHC, resulting in damage to more than 50 superconducting magnets and a 1-year downtime to the LHC. But even at much lower energies, the continuous, localised impact of particles can create damage after some integration time as the local energy loss per proton is strongly increasing at energies below 1 GeV.

### 1.3.2   *Quench protection, heat load and radiation effects*

Beyond these worst-case failure cases leading to irreversible damage to accelerator equipment, the operation of such high intensity machines also requires mastering the collision debris that exit the collision points at low angles and travel along the vacuum chamber of the machine. Already the loss of as little as $10^{-6}$ to $10^{-7}$ of the total beam intensity in one of the superconducting magnets can, in the case of the LHC, result in the loss of the superconducting state and lead to a magnet quench. While this is a recoverable failure without long-term damage to the equipment, large numbers of quenches have a very detrimental effect on machine efficiency and to the achievable physics output of the machine. Efficiently cleaning these highly energetic particles with a muti-stage collimation system as well as passively protecting the superconducting coils through shielding (the HL-LHC project is using tungsten absorbers integrated onto the beam screens of the most exposed inner triplet quadrupole magnets) contributes to a reliable operation of the machine and helps reducing the long-term radiation dose to the most sensitive components of the accelerator. Beyond the long-term radiation dose to equipment, radiation induced failures can also occur in electronics which is installed close to the beamline of the accelerator. Shielding or relocation strategies have shown beneficial to mitigate these effects, as the development of radiation tolerant electronics is a very resource and time intensive process.

An additional effect that must be considered in high intensity colliders is the formation and effects of electron cloud. In the presence of high intensity proton beams, electrons are inevitably produced inside the beam vacuum chamber. Seed electrons (e.g. from synchrotron radiation) will impact the chamber's walls and eject secondary electrons, which subsequently will be accelerated by the proton bunches. When they in-turn hit the vacuum chamber wall, they have a probability of ejecting more electrons, which creates a continues cycle, and the electron cloud (e-cloud) is created. The parameter of interest is hereby the Secondary Emission Yield (SEY), which is the average number of electrons produced per impact. The e-cloud both affects beam quality and considerably increases the heat load on the cryogenic system, representing a potential intensity limitation and increasing the electricity cost for the cryogenic infrastructure. Experience in the LHC [2] and other high intensity machines has shown that an appropriate coating of the beam screens (through e.g. amorphous carbon coating or alternative techniques such as the laser engineered surface treatment [3]) is a must.

Finally, the high intensity beams also interact with particles that remain in the beam vacuum chamber due to contaminations. Dust particles which are inevitably present in the 27 km long vacuum chamber of the LHC have shown to be prone to ionisation by the circulating high intensity proton beams, allowing them to travel along the electric and magnetic fields of the beam and the surrounding magnets and to occasionally interact with the passing proton bunches. These so-called UFO-events manifest as localised beam losses of very short duration ($10^{-4}$ to few $10^{-3}$ seconds) which affect accelerator operation and can even generate magnet quenches. The UFO rates have found to be particularly large following long shutdown periods which imply the venting of the vacuum sectors and therefore the risk of introducing additional contamination. Long operational periods at increasing beam intensities show however a strong conditioning effect (see Fig. 5), which – along with an optimisation strategy of beam loss thresholds – has allowed reliable operation with only a few magnet quenches throughout the second and third operational run of the LHC.

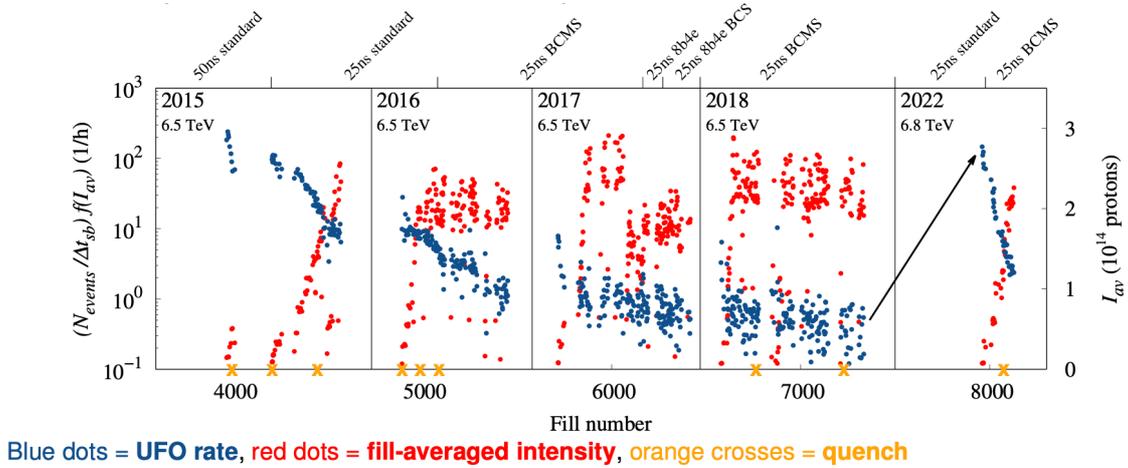

Blue dots = **UFO rate**, red dots = **fill-averaged intensity**, orange crosses = **quench**

**Fig. 5:** UFO rates in the LHC during the second operational run (2015-2018) and the first year of operation of the third run in 2022 (following Long Shutdown 2).

### *1.3.3   Magnet technology and accelerator size*

The achievable collision energy of a collider is directly dependant on the size of the tunnel and the bending field that can be reached in the dipole magnets. For decades, the workhorse of accelerator magnets has been the Nb-Ti technology, which has proven a very robust and comparably cost-efficient material for use on an industrial scale as required for large-scale applications such as Tevatron, RHIC, HERA and most recently the LHC. The HL-LHC upgrade project, as well as future projects such as the Future Circular Collider (FCC) do however require larger magnetic fields for the realisation of their objectives and therefore rely on the use of novel superconducting materials. The High Luminosity project has successfully achieved the industrialisation of $Nb_3Sn$ for its final focusing quadrupole magnets, a technology that is paving the way to magnets featuring up to 16 T peak fields (see Fig. 6). $Nb_3Sn$ has today already become a commodity, with other large-scale projects such as ITER requiring some 500 tons of this material for its superconducting magnets. It comes however at a five times higher cost than Nb-Ti as well as at a much higher fragility. Future colliders may well explore other High Temperature Superconductors (HTS), which are offering even higher magnetic fields but still requiring extensive R&D to achieve the required level of maturity for their use in accelerator magnets.

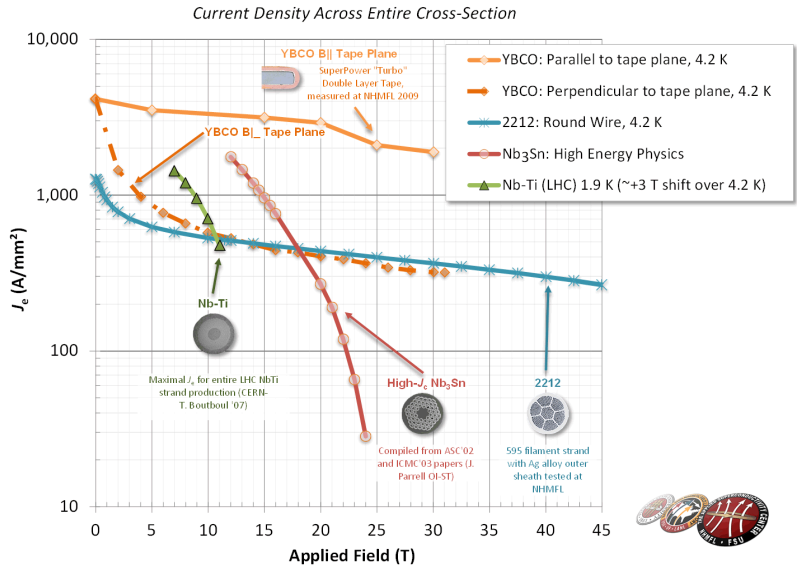

**Fig. 6:** Critical current density and peak field of superconducting materials.

## 2      Present projects and future project options

The latest major HEP projects, including the Tevatron, HERA and the LHC have already been major endeavors, only being possible thanks to a continuous commitment and support of the entire international community to jointly develop the technologies needed for these large-scale scientific instruments. While the next goal of HEP community is clearly defined – the construction of a Higgs-factory – the best machine configuration to achieve this goal is the subject of ongoing discussions. It is however clear that the scope of this project will be ever more ambitious, imply high costs and (decades) long R&D to bring all required technologies to maturity. As an example, the R&D of the Nb3Sn HL-LHC triplet magnets started back in the early 2000s, for an installation in the LHC tunnel planned more than 25 years later in 2029. Due to these increasingly long R&D times, the past trend of continuous increase of center of mass energy over time will not be achieved even with the most ambitious future hadron collider projects such as FCC-hh (see Fig. 7).

Collider projects that are currently in their implementation phase are the Super KEKB and HL-LHC upgrades, while the Electron Ion Collider project has been recently approved in the US. Possible collider projects for the future include ideas for a High Energy LHC (using magnet technologies approaching 16T to double the CM energy), FCC-ee (the present baseline proposal of CERN), CepC or linear collider projects (ILC, CLIC). The ongoing process of the European Strategy aims at converging on the most pragmatic choice for the next generation of the HEP community, aiming at an endorsement at a dedicated European Strategy Council meeting in May 2026.

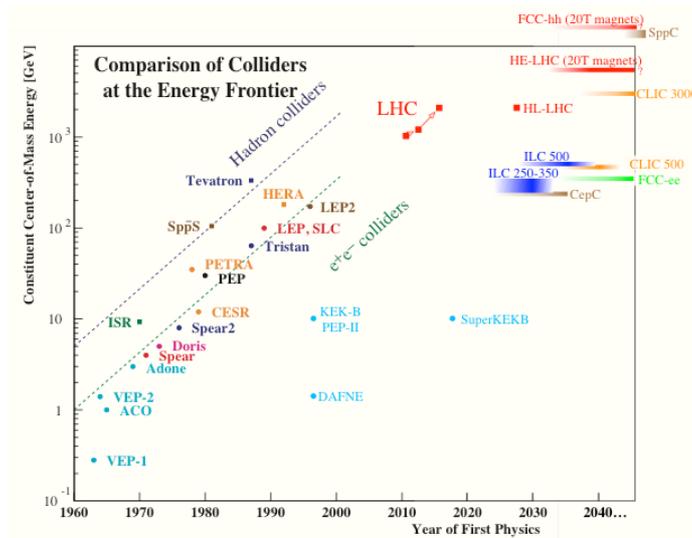

**Fig. 7:** Past, present and future projects at the energy frontier.

## 3   Conclusions

Hadron colliders have evolved into highly sophisticated machines that push the limits of energy, intensity, and luminosity in the pursuit of new physics discoveries. Their performance is fundamentally constrained by competing factors such as beam lifetime, luminosity optimization, and increasing operational complexity. Over time, significant technological advances—particularly in superconducting magnet design, machine protection, and beam control techniques—have enabled these challenges to be effectively managed. Innovations such as luminosity levelling, crab cavities, and advanced collimation systems have proven essential for sustaining high performance and reliability. At the same time, material limitations, radiation effects, risks due to the ever-increasing stored energies continue to demand ongoing research and development. Looking ahead, future projects like the HL-LHC and FCC will rely heavily on these accumulated experiences and emerging technologies. Ultimately, the continued progress of hadron colliders will depend on global collaboration and sustained innovation to unlock deeper insights into fundamental physics.